\documentclass{article}
\usepackage{flushend}
\usepackage{spconf,amsmath,graphicx}
\usepackage{subfigure}
\usepackage{multirow}
\usepackage{longtable}
\usepackage{amsthm}
\usepackage{amssymb}
\usepackage{amsmath}
\usepackage{bm}
\usepackage{mathrsfs}
\usepackage{amsfonts}
\usepackage{longtable}
\usepackage{multirow}
\usepackage{algorithm, algpseudocode}
\usepackage{cite}
\usepackage{amsmath}

\algnewcommand\Input{\item[\hspace{6pt}\textbf{Input:}]}
\algnewcommand\Output{\item[\hspace{6pt}\textbf{Output:}]}
\algnewcommand\OutputVal{\textbf{output} }
\usepackage{tabularx}
\newcolumntype{L}{>{\raggedright\arraybackslash}X}
\usepackage{makecell}
\usepackage{color}
\usepackage{xcolor}

\title{CT-CAPS: Feature Extraction-based Automated Framework for COVID-19 Disease Identification from Chest CT Scans using Capsule Networks}
\name{\makecell{Shahin Heidarian$^1$,
Parnian Afshar$^2$,
Arash Mohammadi$^2$,
Moezedin Javad Rafiee, MD$^3$,\\
Anastasia Oikonomou, MD$^4$,
Konstantinos N. Plataniotis$^5$,
and Farnoosh Naderkhani$^2$}}
\address{$~^1$Department of Electrical and Computer Engineering, Concordia University, Montreal, QC, Canada \\
$~^2$Concordia Institute for Information Systems Engineering, Concordia University, Montreal, Canada\\
$~^3$Department of Medicine and Diagnostic Radiology, McGill University, Montreal, QC, Canada\\
$~^4$Department of Medical Imaging, Sunnybrook Health Sciences Centre, Toronto, Canada\\
$~^5$Department of Electrical and Computer Engineering, University of Toronto, Toronto, Canada}

\begin{document}
\ninept
\maketitle
\begin{abstract}
The global outbreak of the novel corona virus (COVID-19) disease has drastically impacted the world and led to one of the most challenging crisis across the globe since World War II. The early diagnosis and isolation of COVID-19 positive cases are considered as crucial steps towards preventing the spread of the disease and flattening the epidemic curve. Chest Computed Tomography (CT) scan is a highly sensitive, rapid, and accurate diagnostic technique that can complement Reverse Transcription Polymerase Chain Reaction (RT-PCR) test. Recently, deep learning-based models, mostly based on Convolutional Neural Networks (CNN), have shown promising diagnostic results. CNNs, however, are incapable of capturing spatial relations between image instances and require large datasets. Capsule Networks, on the other hand, can capture spatial relations, require smaller datasets, and have considerably fewer parameters. In this paper, a Capsule network framework, referred to as the "CT-CAPS", is presented to automatically extract distinctive features of chest CT scans. These features, which are extracted from the layer before the final capsule layer, are then leveraged to differentiate COVID-19 from Non-COVID cases. The experiments on our in-house dataset of 307 patients show the state-of-the-art performance with the accuracy of $90.8$\%, sensitivity of $94.5$\%, and specificity of $86.0$\%.

\end{abstract}
\textbf{\textit{Index Terms}: COVID-19 Identification, Capsule Networks, Pulmonary Infection, Chest CT Scan, Deep Learning}
%
\section{Introduction} \label{sec:Introduction}
Past months have been sadly tied up with the novel coronavirus (COVID-19) outbreak resulting in long term damages to the society and people's lives. As such, a global action plan has been established to prevent the spread of this highly contagious disease including attempts to find vaccines and early diagnosis of the disease. The latter is highly important for the immediate isolation and treatment of the COVID-19 positive cases. The gold standard diagnostic test for this disease is the Reverse Transcription Polymerase Chain Reaction (RT-PCR), which is associated with a high false negative rate~\cite{Ai2020}. RT-PCR is also a time-consuming test, which may deprive health authorities from the opportunity of early isolation resulting in further spread of the virus. Recent studies have demonstrated the strong capability of Chest Radiographs (CR) and Computed Tomography (CT) scans in providing distinctive patterns associated with the COVID-19 infection such as bilateral prevalence of Ground Glass Opacities (GGO) with peripheral and subpleural distribution and lower lung zone predominance. Halo and reverse halo signs, focal areas of consolidations and interstitial thickening have also been described~\cite{Hani2020,Carotti2020}. COVID-19 lung manifestations, however, are highly overlapped with the Community Acquired Pneumonia (CAP) findings making their identification challenging and complicated even for experienced radiologists and healthcare professionals.

Recently, several research studies have developed deep learning-based approaches to identify COVID-19 disease from chest radiographs, some of which acquired promising performances~\cite{Afshar2020,Wang2020,Islam2020,Altan2020}. Both chest radiograph and CT scan provide a view of the structures inside the body by sending an electromagnetic radiation to the body and capturing the radiation passed through the body on a detector. However, chest radiographs are single 2D images that provide limited details of the lung infections. CT scans, on the other hand, consist of multiple cross-sectional images of the body (slices), which are seen together to create a 3D representation of the body. Unlike chest radiographs, CT scans can provide a comprehensive illustration of the specific structure of the lung abnormalities. Consequently, CT has a higher sensitivity compared to CR especially in the case of COVID-19 detection, where there is a high overlap with other lung infections such as CAP and therefore a fast and accurate diagnosis is imperative. Due to the radiation exposure caused by CT scans and the chance of virus spread during the scanning, CT scan is mainly recommended as diagnostic tools in emergency settings and to assist radiologists evaluating patients with suspected or confirmed COVID-19 infection~\cite{Redmond2020}. However, chest CT is widely used as a primary diagnostic imaging modality in many countries, especially in those where RT-PCR test resources are low. Therefore, several studies have investigated possible deep solutions to identify COVID-19 from 2D and 3D CT images. Analyzing a series of CT scan slices, however, is more challenging compared to a single chest radiograph as we have to review several images and investigate the evidence of infection over multiple slices.

\noindent
\textbf{Contributions:} In this paper, we propose a fully automated framework based on Capsule Networks~\cite{Hinton2018}, referred to as the ``CT-CAPS", to represent each slice of a CT scan by a small feature map to distinguish COVID-19 cases from non-COVID (CAP and normal) cases. The capsule network is selected to address the failure of the commonly used CNN architectures~\cite{Yamashita2018} in recognizing spatial relations between objects in an image and thus eliminate the need for large labeled datasets to reach a satisfying result. Large datasets are difficult to obtain in the case of a novel and barely recognized disease such as COVID-19. Furthermore, CNN architectures consist of millions of parameters making their training process slow and complex. Medical images, in particular COVID-19 related ones, contain distinctive spatial relations between image instances. For example, there is a distinctive distribution of imaging findings in COVID-19 infection on CT scans, which makes finding spatial relations in the image highly important. Moreover, the superiority of Capsule Networks over their counterparts is demonstrated in some of our previous research studies based on various imaging modalities such as magnetic resonance imaging (MRI) and chest radiographs (CR)~\cite{Afshar2020,Afshar2020b,Afshar2020c,Afshar2020d}.

In this study, a Capsule Network-based feature extractor is presented to detect specific characteristics of CT slices, followed by a Max Pooling Layer to convert slice-level feature maps into patient-level ones. Finally, a stack of fully connected layers are added to make the final decision. The CT-CAPS is trained on a dataset of CT slices labeled by three experienced radiologists~\cite{Afshar2020a} to determine slices demonstrating infection and slices without an evidence of infection. The main focus of our labeling process was to detect distinctive and obvious abnormalities in a CT scan not struggling to identify minor findings of questionable infection. Therefore, CT-CAPS framework benefits from a fast and timely labeling process, which is highly valuable when we are facing an early emergence of a new type of data. The experimental results on our in-house dataset of $307$ patients~\cite{Afshar2020a} demonstrate the capability of CT-CAPS model to be trained on such a coarsely labeled dataset achieving the accuracy of $90.8\%$, sensitivity of $94.5\%$, specificity of $86.0\%$, and Area Under the ROC Curve (AUC) of $0.93$ using the cut-off probability of $0.6$. To improve on the explanability of the proposed CT-CAPS, we incorporated the Grad-CAM localization mapping approach~\cite{Selvaraju2016} to determine the lung regions contributing the most to the final decision.
The $\%95$ Confidence Interval (CI) values are also calculated for all performance matrics using the methods explained in~\cite{Brown2001,Hanley1982}. The trained CT-CAPS model and the testing code are available publicly for open access at https://github.com/ShahinSHH/CT-CAPS.

\section{Related Works} \label{sec:method}
Existing identification methods based on chest CT scans are generally divided into slice-level and patient-level methods. These works can further be classified into segmentation-based or feature extraction-based approaches. Segmentation-based methods~\cite{Nguyen2020,Shi2020,Zhang2020} aim to train a model on a large dataset of annotated lung lesions to detect regions of infection and determine the disease severity and type. Although the lung segmentation task has been well-studied~\cite{Hofmanninger2020}, infection segmentation requires an extensive collaboration with radiologists to perform the sophisticated infection and abnormality annotation task making the training process too complicated and time-consuming. Moreover, in some cases~\cite{Shi2020}, the overall performance is low for scenarios with mild lung infections. As an example of segmentation-based methods, Nguyen \textit{et al.}~\cite{Nguyen2020} used a semi-supervised method based on pre-trained existing segmentation models to detect lung infected regions to be incorporated into a CNN-based classifier via an attention mechanism to increase the classification accuracy. The underlying approach reached the overall accuracy of $88\%$ in the binary slice-level classification (COVID-19 and normal cases). In another study, Shi \textit{et al.}~\cite{Shi2020} proposed a model, which extracts handcrafted radiomics features from the segmented lung and infected regions followed by a feature selection mechanism to feed multi-stage random forest classifiers to classify patients into four groups based on their infection size obtained from the first step. Then a random forest model is trained for each group as the final classifier achieving the accuracy of $87.0\%$, sensitivity of $90.7\%$, and specificity of $83.3\%$. As another example, Zhang \textit{et al.}~\cite{Zhang2020} proposed a two-stage method consisting of a sophisticated lung infection segmentation model, trained on a manually annotated chest CT scan dataset, followed by a 3D ResNet18 classification model~\cite{Hara2017} to classify patients into COVID-19, CAP, and normal cases. Their proposed method achieves the overall accuracy of $92.49\%$.

With regard to the feature extraction-based approaches, different frameworks have been recently introduced, commonly utilizing a CNN-based model. Such methods either use a 3D CNN to analyze the whole CT volume in a single stage, or apply 2D CNNs on CT slices and aggregate slice-level results via an aggregation mechanism. As an example, Wang \textit{et al.}~\cite{Wang2020a} fed a 3D CNN-based classifier with lung regions, segmented by a pre-trained U-Net~\cite{Ronneberger2015}, achieving accuracy of $90.1\%$, sensitivity of $84.0\%$, and specificity of $98.2\%$ on a dataset containing only COVID-19 and normal cases.
Hu \textit{et al.}~\cite{Hu2020} extended patient-level labels into slice-level and used the same label for all slices in a CT scan to train a deep model, utilizing the intermediate CNN layers to obtain classification features. These features are then combined to make the final decision. Their proposed method achieved the overall accuracy of $87.4\%$ in the three-way classification. It is worth mentioning that using patient-level labels for all slices in a CT scan is not reasonable and will add errors into the system as each volume of CT scan contains many slices without any visible infection area. In another study, Li \textit{et al.}~\cite{Li2020} proposed a fully automated framework using a ResNet50 as the backbone to extract slice-level features followed by max pooling and fully connected dense layers for the ultimate goal of patient-level classification achieving the sensitivities of $90\%$, $87\%$, and $94\%$ for COVID-19, CAP, and normal cases, respectively. The aforementioned methods either require a carefully annotated data to segment regions of infection, or extend patient-level labels to all slices resulting in an unexplainable and potentially lower results. Moreover, some of the aforementioned works have only proposed slice-level classifiers, which makes such methods partially automated.

\section{CT-CAPS Framework} \label{sec:method}
\begin{figure}[t!]
\centering
\includegraphics[width=0.4\textwidth]{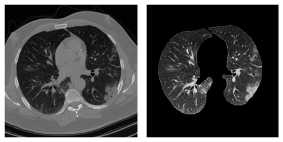}
\caption{\footnotesize The segmented lung tissue using R231CovidWeb model. \label{fig:segment}}
\end{figure}
In this section, the CT-CAPS framework is introduced in detail. First, the in-house dataset of chest CT scans and the associated labeling and preprocessing steps used in this study are briefly described. Next, the detailed description of the model structure and processing pipeline is presented.

\subsection{COVID-CT-MD Dataset} \label{sec:data}
\begin{figure*}[t!]
\centering
\includegraphics[width=0.8\textwidth]{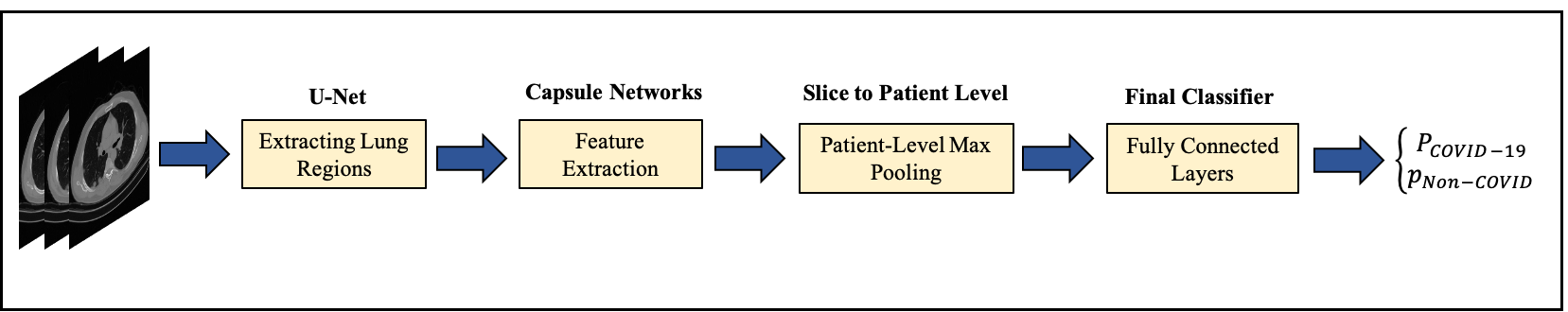}
\caption{\footnotesize The proposed CT-CAPS framework to identify COVID-19 and non-COVID cases from chest CT scans. \label{fig:pipe}}
\end{figure*}
We used our recently released dataset of volumetric chest CT scans referred to as the ``COVID-CT-MD''~\cite{Afshar2020a}, which includes $171$ patients positive for COVID-19 infection, $60$ patients with CAP, and $76$ normal patients. A subset of $55$ COVID-19, and $25$ CAP cases in the COVID-CT-MD are analyzed by three experienced radiologist to detect slices with a distinctive evidence of infection. As mentioned earlier, there exist many slices without evidence of infection in a chest CT scan and commonly only a proportion of the lung is involved in the infection. The labeling process aims to specify slices with distinctive disease manifestations in a timely manner rather than those with minimal findings. The labeled subset of the data contains $4,993$ slices demonstrating infection and $18,416$ slices without evidence of infection. This labeled data is then randomly divided into three groups, including $60\%$, $10\%$, and $30\%$ independent parts of the data, to train, validate, and test the feature extraction stage of the CT-CAPS framework. The remaining data is split with the same proportion and used along with the labeled data to train and evaluate the subsequent fully connected layer as the patient-level classifier. This research work is performed based on the policy certification number 30013394 of Ethical acceptability for secondary use of medical data approved by Concordia University. The COVID-CT-MD dataset is available online through Figshare~\footnote{https://figshare.com/s/c20215f3d42c98f09ad0}.
For more information about the imaging settings and detailed description of the dataset, please refer to Reference~\cite{Afshar2020a}.

\subsection{Lung Segmentation} \label{sec:seg}
In order to remove the unimportant components and artifacts existing in a CT scan, we utilized a pre-trained U-Net-based lung region segmentation model~\cite{Hofmanninger2020} referred to as the ``U-net (R231CovidWeb)'', which has been fine-tuned specifically on the COVID-19 images. A sample of lung region extracted by this model is illustrated in Fig.~\ref{fig:segment}. It is worth mentioning that unlike segmenting infected regions, lung region segmentation is a well-studied topic and highly efficient models have been introduced so far.
The input of the R231CovidWeb model is a CT scan with the original slice size of $512\times512$. The model returns the extracted lung tissues, which will further go through some normalization and resizing steps. More specifically, the output images will be normalized between $0$ and $1$ to help the generalizability and effective convergence of the model. Following the literature~\cite{Zhang2020, Hu2020}, we down-sampled the output images from the original $512\times512$ size to $256\times256$ to reduce the complexity and memory requirements with negligible loss of information. Finally, slices without visible lung tissues are excluded and the remaining ones are saved to be used in the CT-CAPS framework.

\subsection{CT-CAPS Architecture}
\begin{figure*}[t!]
\centering
\includegraphics[width=0.8\textwidth]{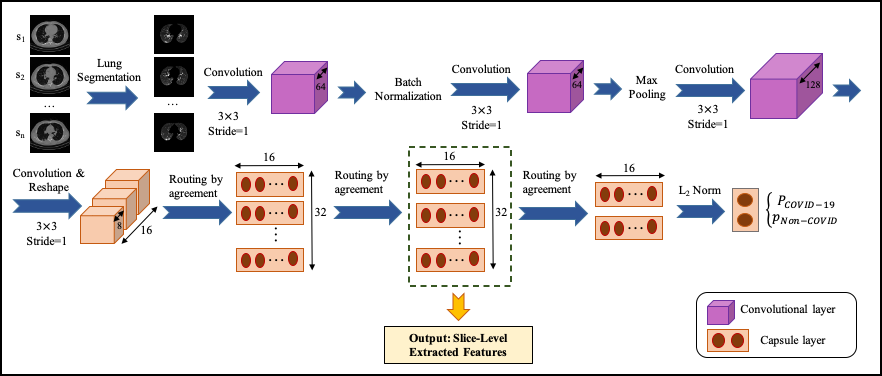}
\vspace{-.1in}
\caption{\footnotesize The Capsule Network model used to extract slice-level features. \label{fig:model}}
\end{figure*}
The CT-CAPS' pipeline is illustrated in Fig.~\ref{fig:pipe}. The first stage of the CT-CAPS architecture consists of a 2D Capsule Network, which aims to classify 2D CT slices into COVID-19 and non-COVID images. This network is initialized by a stack of four convolutional layers along with a batch-normalization and one max pooling layer as shown in Fig.~\ref{fig:model}. The last convolutional layer is then used to feed the subsequent Capsule Layers to extract deeper and smaller feature maps by the routing by agreement process defined in~\cite{Hinton2018}. Two more capsule layers are subsequently added to the model, where the amplitude of the last one represents the probability of the input image belonging to each target class. In the next step, we aggregate slice-level features extracted by intermediate layers of the described network to move on to the patient-level domain. In this regard, the capsule layer before the last one is used as the representative feature map of the slices. Experimental results (presented later in Section~\ref{sec:result}) demonstrate the ability of this feature map to efficiently distinguish between COVID-19 and non-COVID images.

A max pooling layer is then applied on the feature maps associated with slices of a patient and the result is fed to a stack of four fully connected layers with the size of $256$, $128$, $32$, and $2$ respectively. Using this feature selection and max pooling, each 3D volume CT scan is represented by a small $32\times16$ matrix. In addition, we modified the loss function to compensate the relatively imbalanced training dataset. We used a weighted version of the loss function, such that a higher penalty rate is given to the less frequent class, which is COVID-19 in this case. For the fully connected layers, however, the class weights are equal. The loss function of the Capsule Network model is modified as follows
\begin{equation}
\label{loss-function}
loss = \frac{N^{+}}{N^{+}+N^{-}}\times loss^{-} + \frac{N^{-}}{N^{+}+N^{-}}\times loss^{+},
\end{equation}
where $N^{+}$ represents the number of COVID-19 samples, $N^{-}$ is the number of non-COVID samples, $loss^{+} $ signifies the loss value associated with COVID-19 samples, and $loss^{-}$ is the loss value associated with non-COVID samples.

\section{Experimental Results} \label{sec:result}
\begin{table*}[t!]
\centering
\label{tab:res}
\caption{\footnotesize CT-CAPS's Patient-Level Classification Results. Values in parenthesis represent the $95\%$ confidence interval.}
\begin{tabular}{|c|c|c|c|c|}
\hline
\textbf{Performance} & \textbf{\footnotesize{CT-CAPS}} & \textbf{\makecell{\footnotesize{CT-CAPS}\\\footnotesize{(no lung)}}} & \textbf{\footnotesize{CT-CNN}} & \textbf{\footnotesize{CT-Res50}} \\[1ex]
\hline
\textbf{Accuracy(\%)} & \makecell{\textbf{89.8}\\$(89.2,90.4)$} & \makecell{$82.6$\\$(81.8,83.4)$} & \makecell{$78.6$\\$(77.8,79.4)$} & \makecell{$81.6$\\$(80.8,82.4)$} \\
\hline
\textbf{Sensitivity(\%)}& \makecell{$94.5$\\$(93.7,95.3)$} & \makecell{$87.3$\\$(82.4,85.0)$} & \makecell{$87.3$\\$(86.1,88.5)$} & \makecell{\textbf{96.4}\\$(95.7,97.1)$} \\
\hline
\textbf{Specificity(\%)} & \makecell{\textbf{83.7}\\$(82.1,85.4)$} & \makecell{$76.6$\\$(74.7,78.5)$} & \makecell{$67.4$\\$(65.3,69.5)$} & \makecell{$62.8$\\$(60.6,65.0)$}\\
\hline
\textbf{AUC} & \makecell{\textbf{0.93}\\$(0.88,0.98)$} & \makecell{$0.86$\\$(0.78,0.93)$} & \makecell{$0.79$\\$(0.70,0.88)$} & \makecell{$0.82$\\$(0.74,0.90)$} \\
\hline
\textbf{\# Params.} & \textbf{0.5M} & \textbf{0.5M} & $243.9M$ &
$24M$ \\
\hline
\end{tabular}
\end{table*}
The feature extraction part of the CT-CAPS is trained on a dataset of $55$ COVID-19, $25$ CAP, and all $78$ normal cases. The Adam optimizer with the initial learning rate of $1e-4$, batch size of $16$, and $100$ epochs is used in the corresponding training step. For the fully connected patient-level classifier, the initial learning rate of $1e-3$, and $500$ epochs are used. In each stage, the model with the lowest loss value on the validation set is considered as the final model for evaluation. The evaluation results on the COVID-CT-MD dataset~\cite{Afshar2020a} are presented in Table~\ref{tab:res}. We compared the CT-CAPS with its duplicate but using the whole CT images without extracting the lung tissues. We also replaced the Capsule Layers in another experiment with two fully connected layers with the size of $128$, while the rest of the architecture and parameters are kept the same. We then took the fully connected dense layer before the last layer as the new feature map to make a CNN-based alternative model for the comparison.

In a similar experiment, Resnet50, which is the backbone of many similar works such as Reference~\cite{Li2020} is used in the feature extraction stage. In this case, the fully connected layer with $2,048$ neurons before the last layer is taken as the feature map. The comparison results are presented in Table~\ref{tab:res}. The CT-CAPS achieves the accuracy of $89.8\%$, high sensitivity of $94.5\%$, specificity of $83.7\%$, and Area Under the ROC Curve (AUC) of $0.93$ using the default probability threshold of $0.5$. It is worth mentioning that various cut-off probabilities can be used based on the radiologist's preference. A higher cut-off probability indicates the higher importance of the specificity, while the lower value indicates the higher importance of the sensitivity. In other words, radiologists can adjust this threshold to change the range of required confidence to consider a patient as a COVID-19 positive case or a non-COVID case. Experimental results show that increasing the probability threshold from $0.5$ to $0.6$ improves the accuracy to $90.8\%$, and the specificity to $86.0\%$ while the sensitivity remains the same. Table~\ref{tab:cutoff} presents performance of the proposed CT-CAPS using different cut-off probabilities.

\begin{figure}[t!]
\centering
\includegraphics[width=0.4\textwidth]{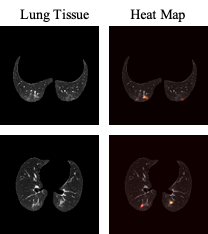}
\label{fig:heatmap}
\vspace{-.1in}
\caption{\footnotesize The heat-maps created by the GRAD-CAM approach from the last convolutional layer of the CT-CAPS framework for two sample images with COVID-19-related evidences of infection.}
\vspace{-.2in}
\end{figure}

\begin{table}[t!]
\centering
\label{tab:cutoff}
\caption{\footnotesize Performance of the CT-CAPS using different cut-off probabilities. Values in parenthesis represent the $95\%$ confidence interval.}
\begin{tabular}{|c|c|c|c|}
\hline
\makecell{\textbf{Cut-off}\\\textbf{Probability}} & \textbf{Accuracy(\%)} & \textbf{Sensitivity(\%)} & \textbf{Specificity(\%)} \\[1ex]
\hline
\textbf{0.3} & \makecell{$86.7$\\$(86.0,87.4)$} & \makecell{$94.5$\\$(93.7,95.3)$} & \makecell{$76.7$\\$(74.8,78.6)$} \\
\hline
\textbf{0.4}& \makecell{$88.8$\\$(88.2,89.4)$} & \makecell{$94.5\%$\\$(93.7,95.3)$} & \makecell{$81.4\%$\\$(79.6,83.2)$} \\
\hline
\textbf{0.5} & \makecell{$89.8$\\$(89.2,90.4)$} & \makecell{$94.5\%$\\$(93.7,95.3)$} & \makecell{$83.7\%$\\$(82.0,85.4)$} \\
\hline
\textbf{0.6} & \makecell{$90.8$\\$(90.2,91.4)$} & \makecell{$94.5\%$\\$(93.7,95.3)$} & \makecell{$86.0\%$\\$(84.4,87.6)$} \\
\hline
\textbf{0.7} & \makecell{$89.8$\\$(89.2,90.4)$} & \makecell{$90.9\%$\\$(89.9,92,0)$} & \makecell{$88.4\%$\\$(86.9,89.9)$} \\
\hline
\end{tabular}
\vspace{-.2in}
\end{table}
In addition to the aforementioned numerical results, we incorporated the GRAD-CAM localization mapping method to visualize the distinctive patterns in a chest CT scan recognized by the last convolutional layer of the CT-CAPS. Fig.~\ref{fig:heatmap} illustrates the recognized abnormal regions for two lung samples containing small evidences of COVID-19 infection. In these two examples, we can observe that the model correctly identified the regions of infection, which had the highest contribution to final decision.

\vspace{-.15in}
\section{Conclusion}  \label{sec:con}
\vspace{-.1in}
In this paper, we proposed a fully automated Capsule network-based framework, referred to as the ``CT-CAPS", which extracts distinctive features of chest CT scans via Capsule Networks to identify COVID-19 cases in a coarsely-labeled dataset of COVID-19, CAP, and normal cases. The experimental results indicate the capability of the CT-CAPS to automatically analyze volumetric chest CT scans and distinguish different cases with the accuracy of $90.8\%$, high sensitivity of $94.5\%$, and specificity of $86.0\%$, using far less number of parameters and less sophisticated labeling process compared to its existing counterparts. We also showed that Capsules in the layer before the last one can be a proper compact feature representative of CT scans. Moreover, the benefits of extracting lung tissues in the CT-CAPS framework, and the flexibility of the model to be adjusted based on radiologists' preferences to achieve desired results have been demonstrated by the experimental results. As a final note, we will continue to further improve and validate our proposed method upon receiving new datasets from our collaborators in medical centers to modify our model based on a multi-center dataset.

\bibliographystyle{IEEEbib}
\bibliography{refs}

\end{document}